\documentclass[preprint,12pt]{elsarticle}

\usepackage{amssymb}

\usepackage{color} 
\begin{document}

\begin{frontmatter}



\title{Intermediate state switching dynamics in magnetic double layer
nanopillars grown by molecular beam epitaxy}


\author [1,3]{N. M\"usgens}
\author [1,3] {T. Maassen\fnref{fn1}}

\author [1,3]{E. Maynicke}
\author[1,3]{S. Rizwan Ali}

\author [2,3]{A. Heiss}
\author[2,3]{R. Ghadimi}
\author[2,3]{J. Mayer}

\author[1,3]{G. G\"untherodt}
\author[1,3]{B. Beschoten\corref{cor1}}
\ead{bernd.beschoten@physik.rwth-aachen.de}

\date{\today}
\fntext[fn1] {Current address: Physics of Nanodevices, Zernike
Institute for Advanced Materials, University of Groningen, The
Netherlands}

\cortext[cor1]{Corresponding author}

\address [1]{II. Institute of Physics, RWTH Aachen University,
52056 Aachen, Germany}
\address [2] {Central Facility for Electron Microscopy, RWTH Aachen
University, 52056 Aachen, Germany}
\address [3] {JARA-Fundamentals of Future
Information Technology, J\"ulich-Aachen}

\begin{abstract}

We observe a stable intermediate resistance switching state in the
current perpendicular to plane geometry for all Co/Cu/Co double
layer nanopillar junctions grown by molecular beam epitaxy. This
novel state has a resistance between the resistances of the parallel
and antiparallel alignment of both Co-layer magnetizations. The
state, which originates from an additional in-plane magnetic easy
axis, can be reached by spin transfer torque switching or by an
external magnetic field. In addition to spin torque-induced coherent
small-angle spin wave modes we observe a broad microwave emission
spectrum. The latter is attributed to incoherent magnetic
excitations that lead to a switching between the intermediate state
and the parallel or antiparallel alignment of both ferromagnetic
layers. We conclude that the additional magnetic easy axis
suppresses a stable trajectory of coherent large-angle precession,
which is not observed in our samples.

\end{abstract}


\end{frontmatter}


\section{Introduction}
A spin-polarized current passing through a ferromagnetic layer
can exert a spin transfer torque on its magnetization
\cite{JoMaMM159_Slonczewski1996_Current-DrivenExcitationofMagneticMultilayers,
PRB54_Berger1996_EmissionofSpinWavesbyaMagneticMultilayerTraversedbyaCurrent,
PRL80_Tsoi1998_ExcitationofaMagneticMultilayerbyanElectricCurrent,
S285_Myers1999_Current-InducedSwitchingofDomainsinMagneticMultilayerDevices,
PRB66_Stiles2002_AnatomyofSpin-transferTorque}. The torque is of
particular interest, since it opens the way to an efficient control of the
magnetic moments of a nanomagnet in applications such as nonvolatile magnetic
memories and tunable microwave oscillators. Most previous work has been
focused on polycrystalline nanopillars consisting of ferromagnet/normal metal/ferromagnet
(FM/NM/FM) layered systems with unequal FM layer thicknesses.
The thicker FM layer is used to spin-polarize the current when
flowing in the perpendicular-to-plane direction. The second NM/FM interface
may exhibit a net spin torque on the thinner FM due to the absorption of a
transverse spin component. Depending on both the current and the external
 magnetic field the spin torque can either switch the thin FM layer
 hysteretically between the parallel (P) and antiparallel (AP) alignment
 of both magnetizations \cite{PRL84_Katine2000_Current-DrivenMagnetizationReversalandSpin-WaveExcitationsinCoCuCoPillars} or excite the thin layer in a variety of distinguishable dynamical modes \cite{N425_Kiselev2003_MicrowaveOscillationsofaNanomagnetDrivenbyaSpin-PolarizedCurrent, NM3_Lee2004_ExcitationsofIncoherentSpin-WavesDuetoSpin-TransferTorque,
 PRL93_Kiselev2004_Current-InducedNanomagnetDynamicsforMagneticFieldsPerpendiculartotheSamplePlane}. In epitaxial bilayer samples, however, magneto-crystalline
anisotropy can reveal an additional anisotropy axis resulting in
two-step magnetization switching effects
\cite{APL89_Dassow2006_NormalandInverseCurrent-inducedMagnetizationSwitchinginaSingleNanopillar,
PRB76_Lehndorff2007_AsymmetricSpin-transferTorqueinSingle-crystallineFe-Ag-FeNanopillars}.

In this work, we study Co/Cu/Co-nanopillars grown by molecular beam
epitaxy (MBE) using quasi-static and high-frequency resistance
measurements at room temperature. The goal is to establish a
field-current phase diagram of the differential resistance and the
microwave emission of theses junctions. We observe a stable
intermediate resistance state (IS) in between the P and the AP
alignment of both ferromagnetic layers. We discuss the origin of the
IS and its effect on the magnetization dynamics of the thin Co
layer. We find that the novel IS suppresses coherent large-angle
precessional magnetization dynamics favoring incoherent
magnetization excitations.

\section{Materials and methods}

Samples have been fabricated using focused ion-beam assisted
nanostencil masks
\cite{JoAP101_Ozyilmaz2007_Focused-Ion-BeamMillingBasedNanostencilMaskFabricationforSpinTransferTorqueStudies, JoPDAP41_Muesgens2008_Current-InducedMagnetizationDynamicsinSingleandDoubleLayerMagneticNanopillarsGrownbyMolecularBeamEpitaxy}.
The thin film stack of Co(3 nm)/Cu/Co(15 nm) is deposited by
MBE in prefabricated undercut templates on top of a sputtered Pt
bottom electrode (BE) (Fig.~1(a), inset). The presented experimental
data have been obtained from a nanopillar with an elliptical
cross section of $30\times 90~\mathrm{nm}^2$ and a 15~nm thick Cu
spacer. While we focus in this paper on the results obtained on a
single sample, the existence of the intermediate resistance state
is independent of the Cu spacer thickness ($10-25$ nm) and the cross
sectional area ($30\times 60~\mathrm{nm}^2$ to $50\times 150~\mathrm{nm}^2$).
The device is connected to microwave probes and a DC~current
is injected via a bias-tee, along with a $100~\mathrm{\mu A}$ AC~current
($f=1132~\mathrm{Hz}$). The high frequency components of the
resistance changes due to magnetization dynamics are analyzed
using a spectrum analyzer with a bandwidth between 
0.1 and $20~\mathrm{GHz}$. All measurements are performed at 300~K
with an in-plane magnetic field aligned along the easy axis due to
the shape anisotropy. For each microwave power spectrum $\Delta
P(I,H,f)$ we have subtracted a background spectrum. We define
electrons flowing from the thin to the thick Co layer as positive
current.

\begin{figure}
\includegraphics{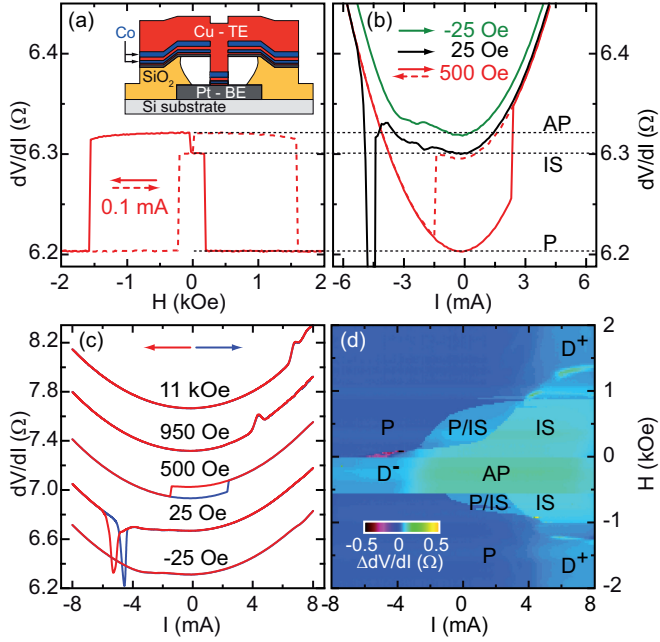}
\caption{\label{fig1} (color online) (a) Differential resistance
$dV/dI$ vs.~magnetic field $H$ at $I = 0.1$~mA. TE and BE denote top
and bottom electrode, respectively.
Inset: Schematics of the nanopillar device. (b) $dV/dI$ vs $I$,
related to the magnetoresistance in (a). (c) $dV/dI$ vs $I$ at
representative magnetic field values for both sweep directions.
Curves are offset for clarity. (d) Variation of $dV/dI$ vs $I$ and
$H$. P, AP, and IS, denote the parallel, antiparallel, and
intermediate resistance state, respectively. D$^±$ are the dynamical
regimes for positive (+) and negative (-) currents.}
\end{figure}

\section{Results}

Fig.~1(a) shows a typical field-driven magnetoresistance (MR)
measurement at small positive current. When $H$ is reduced from $H =
2$~kOe (red solid line) the dipolar coupling between both Co layers
causes the thin layer to switch abruptly already at $H_1=200$~Oe
from the low resistance P state to a higher resistance state. In
contrast to previous studies
\cite{JoAP101_Ozyilmaz2007_Focused-Ion-BeamMillingBasedNanostencilMaskFabricationforSpinTransferTorqueStudies,
APL83_Urazhdin2003_EffectofAntiferromagneticInterlayerCouplingonCurrentassistedMagnetizationSwitching}
this higher resistance state does not correspond to the AP state but
to a novel stable IS with a resistance value lower than the AP
state. The latter state is only reached at $H_2= -40$~Oe, which is
the coercive field of the thick layer. As the negative $H$ is
increased further the thin layer switches at $H_3= -1570$~Oe in one
step to the reversed P configuration. From the switching fields
$H_1$ and $H_3$ we can extract \cite{Rechnung} the coupling strength
between the two FM layers $H_{\rm{coupling}}=885$~Oe and the
coercive field of the thin layer $H_{\rm{C, thin}}=685$~Oe,
consistent with previous studies of Co-based nanopillars
\cite{N425_Kiselev2003_MicrowaveOscillationsofaNanomagnetDrivenbyaSpin-PolarizedCurrent,
APL83_Urazhdin2003_EffectofAntiferromagneticInterlayerCouplingonCurrentassistedMagnetizationSwitching}.

In Fig.~1(b) we depict the differential resistance $dV/dI$ of the
junction as a function of $I$ for representative values of $H$. The
system was first set into the P state at $I=-8~\mathrm{mA}$ and
$H~=~2$~kOe. For $H=500$ Oe we clearly observe current induced
hysteretic switching. In comparison with the field-driven MR
(Fig.~1(a)) it is obvious that the high resistance state is not the
AP state but rather corresponds to the IS. Note that even for
current values up to 15 mA the AP state is not accessible (not
shown). Such a behavior has not been observed for sputtered samples
\cite{PRL84_Katine2000_Current-DrivenMagnetizationReversalandSpin-WaveExcitationsinCoCuCoPillars,
JoAP101_Ozyilmaz2007_Focused-Ion-BeamMillingBasedNanostencilMaskFabricationforSpinTransferTorqueStudies,
APL77_Albert2000_Spin-PolarizedCurrentSwitchingofaCoThinFilmNanomagnet}.
We observe a similar behavior in all our junctions.

For large positive currents and fields we observe the well known
peaks in the differential resistance (in Fig.~1(c)), which indicate
the onset of spin wave generation in the thin layer
\cite{N425_Kiselev2003_MicrowaveOscillationsofaNanomagnetDrivenbyaSpin-PolarizedCurrent}.
At small magnetic fields (see $H=25$~Oe curve in Fig.~1(b) and (c))
pronounced non-hysteretic dips are seen in the $dV/dI$ curves. As
our sample is in the IS and not in the AP state, it will be
interesting to explore the spin wave dynamics in this regime. As the
field polarity is inverted (see, e.g., $H=-25$~Oe in Fig.~1(b)) the
system switches into the AP state. In this state only a small
resistance decrease appears at negative $I$.

The current induced transport is summarized in a transport phase
diagram in Fig.~1(d). The three distinct resistance states (P, IS,
AP) and the expected dynamics regions (D$^±$) are marked. The major
difference compared to previous phase diagrams of sputtered Co/Cu/Co
samples
\cite{APL83_Urazhdin2003_EffectofAntiferromagneticInterlayerCouplingonCurrentassistedMagnetizationSwitching,
PRB72_Sankey2005_MechanismsLimitingtheCoherenceTimeofSpontaneousMagneticOscillationsDrivenbyDCSpin-PolarizedCurrents}
is that our samples can be driven by current only into the IS and
not into the AP state. Otherwise the phase diagrams looks quite
similar.

\begin{figure}
\includegraphics{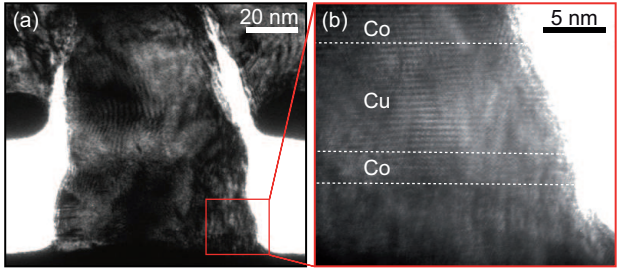}
\caption{\label{fig2} (color online). (a) Transmission electron
microscopy (TEM) image of an MBE grown nanopillar consisting of Pt
(bottom electrode)/Cu (5~nm)/Co (3~nm)/Cu (10~nm)/Co (15~nm)/Cu (top
electrode). (b) High resolution TEM image with element resolved
mapping according to electron energy loss spectroscopy.
}
\end{figure}

To further explore the magnetic configuration of the IS, we have
varied the in-plane orientation of the magnetic field. We find that
the switching in the field-driven MR always occurs via the IS,
independent of the angle between $\vec{H}$ and the magnetic easy
axis of the Co layer shape anisotropy (data not shown). Furthermore
the IS does not change its resistance value near $H=0$. This
demonstrates that the IS is given by a well defined magnetization
configuration of both FM layers and does not originate from a
multidomain configuration or a vortex state. A multidomain
configuration is also not plausible as we observe the IS in every
sample. It is therefore indicative that the magneto-crystalline
anisotropy (MCA) of the FM layers might be related to the IS, which
becomes relevant in epitaxial layers \cite{PRB80_Lehndorff2009_MagnetizationDynamicsinSpinTorqueNano-oscill%
ators-VortexStateversusUniformState}.

To investigate the crystallinity of our MBE grown nanopillars, we
use transmission electron microscopy (TEM). A typical result is
shown in Fig.~2 for a layer sequence of
Cu(5~nm)/Co(3~nm)/Cu(10~nm)/Co(15~nm)/Cu(70~nm) deposited on top of
a Pt bottom electrode. The presented side view image (Fig.~2(a))
demonstrates that the IS does not originate from an artefact of the
extended layer stack on top of the Pt hard mask because theses
layers are separated from the magnetic nanopillar due to its tapered
sidewalls. The high crystallinity of the MBE grown films is visible
in the high resolution TEM image in Fig.~2(b) over the full layer
stack. Although we cannot determine the in-plane crystal orientation
of the layers inside the nanopillar, this result supports the notion
that the MCA of the FM layers is relevant for the magnetic
switching. From the [111] texture of the Co layers \cite{Anmerkung}
we conclude the existence of additional magnetic easy axes in the
layer plane, non-collinear with the easy axis of the shape
anisotropy.

Consequently, the magnetic switching of the thin Co layer from the P
state into the IS is consistent with switching between the easy axis
direction due to the shape anisotropy and an easy axis direction of
the MCA. In order to estimate the resulting angle in between both
Co-magnetization directions in the IS configuration we use for
simplicity the $\sin^2(\varphi)$-dependence of the giant
magnetoresistance effect for symmetrical systems consisting of 2
identical ferromagnetic layers
\cite{JoMaMM247_Slonczcwski2002_CurrentsandTorquesinMetallicMagneticMultilayers}.
For the resistance value of the IS in Fig.~1 (a) this gives an angle
of $\varphi \approx 130^{\circ}$. It is important to emphasize that
the relative resistance
$r_{\rm{IS}}=\frac{R_{\rm{IS}}-R_{\rm{P}}}{R_{\rm{AP}}-R_{\rm{P}}}$
of the IS does vary from junction to junction for otherwise
identical layer stacks. This indicates a variation of the MCA axis
relative to the shape anisotropy direction for different junctions
which is probably related to a random growth orientation on
the sputtered Pt bottom electrode. 

\begin{figure}
\includegraphics{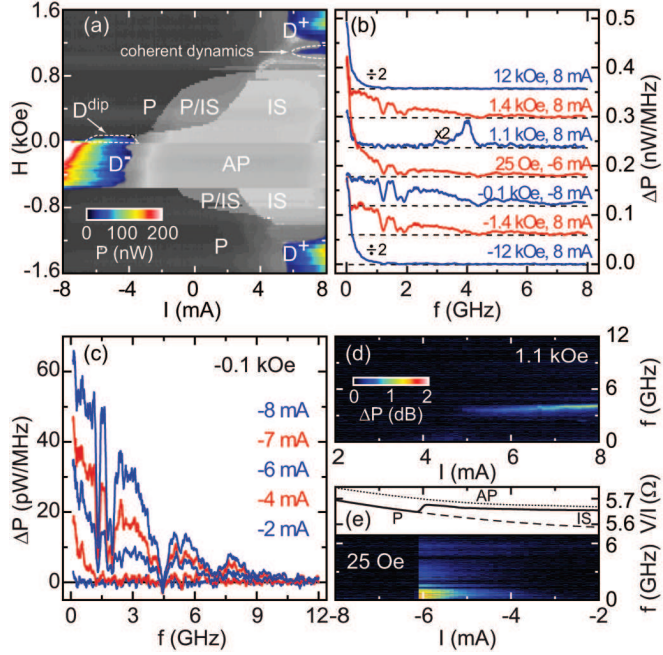}
\caption{\label{fig3} (color online). (a) Integrated microwave power
in a false color plot vs.~$I$ and $H$. $dV/dI$ vs.~$I$ and $H$ (see
Fig.~1(d)) is plotted on a gray scale for comparison. (b) Spectra of
emitted microwave power $\Delta P(I,H,f)$ for different $H$ and $I$
values. (c) $\Delta$P(f) for $H = -0.1$ kOe and $-6$~mA~$\leq $I$
\leq -2$~mA. (d) Current dependence of the microwave power density
$\Delta P(f)$ on a logarithmic scale for $H = 1.1$~kOe. (e)
Microwave power density $\Delta P(f)$ for $H = 25$~Oe (same color
code as in (d)). The black solid line shows $V/I$ vs.~$I$ for the IS
to P transition near $I=-6$~mA; the dotted line corresponds to the
AP state and the dashed line indicates the P state of the junction.}
\end{figure}

We next focus on the spin torque-induced spin wave dynamics. In
Fig.~3 we present frequency domain studies of microwave emission
detected near the current induced magnetization instabilities as
previously identified in the $dV/dI$ vs.~$I$ curves in
Fig.~1(b),(c). The correlation between the differential resistance
changes and the integrated microwave power is displayed in the
microwave emission phase diagram (Fig.~3(a)). It shows different
regimes of excitations depending on both current and magnetic field
polarity \cite{PRB70_Xiao2004_BoltzmannTestofSlonczewskisTheoryofSpin-transferTor%
que}. On the one hand, we observe microwave emission at negative
currents in both the dip regime D$^{dip}$ for small positive fields
(Fig.~1(c)) and the D$^-$-regime. On the other hand, we detect
magnetic excitations at large positive currents and magnetic fields
$|H|>1.2$~kOe (D$^+$-regimes). These excitations are linked to peaks
in $dV/dI$ (Fig.~1(c)), i.e.~to an increase in junction resistance.

In Fig.~3(c) we depict microwave spectra for various negative
currents at a small magnetic field of $-0.1$~kOe (D$^-$-regime). For
currents below the emission threshold of $-2$~mA, we detect
microwave emission spectra extending from 100~MHz up to 9~GHz. Their
overall amplitude $\Delta P$ increases with increasing negative
current values (Fig.~3(c)), but surprisingly, there is no frequency
shift neither with current nor with magnetic field. Note that the
sharp minima in $\Delta P$ observed at several frequencies are
visible in all measured spectra independent of current. They are not
related to the sample but are rather due to absorption in the high
frequency electrical circuit.

Similarly broad emission spectra are also observed in the
D$^+$-regimes (see $H=\pm1.4$~kOe, $I=8$~mA in Fig.~3(b)) as well as
in the dip regime D$^{dip}$ (see $H=25$~Oe emission data in
Fig.~3(b) and in Fig.~3(e) with a continuum at low frequencies). In
the latter case the microwave emission abruptly disappears near
$I=-6$~mA when the magnetization configuration switches from the IS
into the P state. This switching is reflected in the corresponding
change in resistance $V/I$ vs.~$I$ included in Fig.~3(e). We
interpret these broad microwave emission spectra (Fig.~3(b,c,e)) as
a result of incoherent spin-wave excitations due to spin transfer
torque that lead to a switching
\cite{PRL91_Urazhdin2003_Current-DrivenMagneticExcitationsinPermalloy-BasedMultilayerNanopillars,
PRL91_Fabian2003_Current-InducedTwo-LevelFluctuationsinPseudo-Spin-ValveCo-Cu-CoNanostructures}
between the intermediate state IS and the P or AP state of both
ferromagnetic layers. The incoherence is attributed to
inhomogeneities in local fields giving rise to distributions of
local precession frequencies \cite{NM3_Lee2004_ExcitationsofIncoherentSpin-WavesDuetoSpin-TransferTorq%
ue}. Hence, we conclude that the additional easy axis due to the MCA
of the thin Co layer suppresses a stable trajectory, i.e.~a well
defined mode frequency of the coherent large-angle dynamics, which
was previously found in polycrystalline samples
\cite{PRB72_Sankey2005_MechanismsLimitingtheCoherenceTimeofSpontaneousMagneticOscillationsDrivenbyDCSpin-PolarizedCurrents,
N425_Kiselev2003_MicrowaveOscillationsofaNanomagnetDrivenbyaSpin-PolarizedCurrent}.
Note that the observed incoherent magnetization dynamics has not
been reported in frequency domain studies on single-crystalline
nanopillars \cite{MiTo_Lehndorff2008}.

In addition to the incoherent magnetization dynamics we also observe
coherent small angle precessional dynamics of the thin Co layer in a
small range of positive magnetic fields for positive currents
slightly above the hysteretic switching regime (see $H=1.1$~kOe,
5~mA$\leq I \leq8$~mA in Fig.~3(a,b,d)). The spectra have a typical
linewidth of $FWHM \approx 300$~MHz. Independent of the magnetic
field we only observe spin wave frequencies which increase with
increasing current magnitude. Such behavior is characteristic of
out-of-plane precessional modes
\cite{TiAP_Stiles2006_Spin-TransferTorqueandDynamics}. In contrast,
we do not observe in-plane precessions which could be identified by
a decrease of their excitation frequency with increasing current
\cite{TiAP_Stiles2006_Spin-TransferTorqueandDynamics}. These
findings support the notion that in-plane precession of the thin
layer magnetization is inhibited by the intermediate state.

For high magnetic fields $|H|>4$~kOe and large positive currents the
microwave emission continuously evolves into a spectrum resembling
1/f noise without any high-frequency peaks. The reason lies in the
chaotic character of the incoherent spin wave excitations
\cite{NM3_Lee2004_ExcitationsofIncoherentSpin-WavesDuetoSpin-TransferTorque}
as seen in Fig.~3(b) for $H=\pm 12$~kOe. This regime is correlated
with an increase in junction resistance (Fig.~1(c)). The so-called
static high resistance state has also been observed previously in
systems without an intermediate resistance state
\cite{PRB72_Kiselev2005_Spin-TransferExcitationsofPermalloyNanopillarsforLargeAppliedCurrents}
and has its origin in an out-of-plane tilted magnetization of the
thin Co layer, where the spin torque and the torque by the external
magnetic field compensate each other
\cite{PRB68_Li2003_MagnetizationDynamicswithaSpin-transferTorque,
PRB70_Xiao2004_BoltzmannTestofSlonczewskisTheoryofSpin-transferTorque,
PRB72_Slavin2005_Current-inducedBistabilityandDynamicRangeofMicrowaveGenerationinMagneticNanostructures}.
The out-of-plane tilted magnetization therefore yields a transition
to a regime of dynamics not affected by the in-plane anisotropies.

\section{Conclusions}

In conclusion, MBE-grown Co/Cu/Co magnetic double layer nanopillar
junctions show a stable intermediate resistance state with a
resistance value between the resistances of the parallel and
antiparallel alignments of both Co layers. It results from magnetic
switching of the thin Co layer into an additional in-plane easy
axis, which is non-collinear with the shape anisotropy axis. We have
shown that the additional easy axis may originate from a
magneto-crystalline anisotropy due to the texture or enhanced
crystallinity of the layer stack. For magnetic fields smaller than
the coercive field of the thin Co layer, we observe hysteretic
switching between the parallel magnetization configuration and the
intermediate resistance state. However, no current induced
transitions into the AP state have been observed. Although the
current induced resistance changes agree qualitatively with dipolar
coupled double magnetic layer nanopillars without an intermediate
state, no coherent large-angle precessional motion of the thin-layer
magnetization has been observed. Instead we detect broad microwave
emission spectra, which we attribute to incoherent excitations that
lead to switching between the intermediate state and the parallel or
antiparallel magnetization configurations.

We acknowledge useful discussions with B. \"Ozyilmaz. This work was
supported by DFG through SPP 1133 and by HGF.

\label{}





\bibliographystyle{model1a-num-names}



\end{document}